\documentclass[pra,twocolumn]{revtex4}
\usepackage{amssymb,epsf,amsmath}

\newcommand{\nn}{\nonumber}
\newcommand{\dd}{{\rm d}}
\newcommand{\e}[1]{\,{\rm e}^{#1}\,}

\newcommand\C{\mathbb{C}}
\newcommand\Hh{\mathcal{H}}
\newcommand\Ham{\mathbb{H}}

\newcommand\E{\mathcal{E}}

\def\Tr{{\operatorname{Tr\,}}}

\def\an#1{a_{#1}^{\phantom{\dagger}}}
\def\ad#1{a_{#1}^\dagger}
\newcommand\id{{\mathbb I}}
\newcommand\N{{\mathbb N}}
\newcommand\eps\varepsilon

\begin{document}

\title[Yrast Line of a Rapidly Rotating Bose Gas]{The Yrast Line of a
  Rapidly Rotating Bose Gas: Gross-Pitaevskii Regime}

\author{Elliott H. Lieb}
\affiliation{Departments of Mathematics and Physics, Princeton University, Princeton, NJ 08544, USA}
\email{lieb@princeton.edu}

\author{Robert Seiringer}
\affiliation{Department of Physics, Princeton University, Princeton, NJ 08544, USA}
\email{rseiring@princeton.edu}

\author{Jakob Yngvason}
\affiliation{Fakult\"at f\"ur Physik, Universit\"at Wien, and 
Erwin Schr\"odinger Institute for Mathematical Physics, 1090 Vienna, Austria}
\email{jakob.yngvason@univie.ac.at}

\begin{abstract}
  We consider an ultracold rotating Bose gas in a harmonic trap close
  to the critical angular velocity so that the system can be
  considered to be confined to the lowest Landau level. With this
  assumption we prove that the Gross-Pitaevskii energy functional
  accurately describes the ground state energy of the corresponding
  $N$-body Hamiltonian with contact interaction provided the total
  angular momentum $L$ is much less than $N^2$. 
While the Gross-Pitaevskii energy is always an
obvious variational upper bound to the ground state energy, a more refined
analysis is needed to establish it as an exact lower bound.
We also
  discuss the question of Bose-Einstein condensation in the parameter
  range considered. Coherent states together with inequalities in
  spaces of analytic functions are the main technical tools.
\end{abstract}

\maketitle

\section{Introduction}
A Bose gas rotating in a harmonic trap has a critical rotation
speed above which the trap cannot confine it against centrifugal
forces. If the trapping potential equals $\frac m2 (\omega_\perp^2 r^2
+ \omega_3 x_3^2)$,
with $m$ the particle mass and $r=\sqrt{x_1^2+x_2^2}$ the distance
from the axis of
rotation, then the critical angular velocity is $\omega_\perp$. In a
reference frame rotating with angular velocity $\Omega$ the
Hamiltonian for one particle is
\begin{equation}\label{h1}
   \frac{p^2}{2m} + \frac m2 \left( \omega_\perp^2r^2+
\omega_3^2x_3^2\right)  - \Omega L
\end{equation}
with $L$ the component of the angular momentum along the rotation
axis. It is convenient and instructive to complete the square and
write (\ref{h1}) as
\begin{equation}\label{h2}
   \frac{1}{2m} \left( p - A \right)^2 + \frac {m\omega_3^2}{2} x_3^2 +
(\omega_\perp - \Omega) L
\end{equation}
where $A= m\omega_\perp(x_2,-x_1,0)$. In the rapidly rotating case,
where $0<\omega_\perp-\Omega\ll \min\{\omega_\perp,\omega_z\}$, it is
natural to restrict the allowed wave functions to the ground state
space of the first two terms in (\ref{h2}), which we denote by $\Hh$,
and this restriction will be made in this paper.  The space $\Hh$
consists of functions in the lowest Landau level (LLL) for motion in
the plane perpendicular to the axis of rotation, multiplied by a fixed
Gaussian in the $x_3$-direction. Apart from the irrelevant additive
constants $2\hbar\omega_\perp$ and $\hbar\omega_3$, the kinetic energy
in $\Hh$ is simply
\begin{equation}
\omega L
\end{equation}
with $\omega = \omega_\perp - \Omega > 0$. Note that $L$ is 
non-negative for functions in $\Hh$. Its eigenvalues are $0,\hbar,2\hbar,
\dots$.

To characterize the functions in the space $\Hh$, it is natural to
introduce complex notation, $z=x_1+i x_2$. Functions in $\Hh$ are of
the form
\begin{equation}\label{fh}
f(z) \exp\left( - \frac{m\omega_\perp}{2\hbar} |z|^2 -  \frac{m
\omega_3}{2\hbar} |x_3|^2\right )
\end{equation}
with $f$ an analytic function. All the freedom is in $f$ since the Gaussian
is fixed.  If the trapping potential in the $x_3$-direction is not
quadratic, the Gaussian in the $x_3$-variable has to be replaced by
the appropriate ground state wave function.

A fancy way of saying this is that our Hilbert space $\Hh$ consists of
analytic functions on the complex plane $\C$ with inner product given by
\begin{equation}\label{scalarproduct}
\langle \phi|\psi\rangle = \int_{\C} \phi(z)^* \psi(z) \e{-|z|^2} dz\,,
\end{equation}
where $dz$ is short for $dx_1\, dx_2$.
For simplicity we  choose units such that $m= \hbar = \omega_\perp=1$.
The eigenfunction of the angular momentum $L$ corresponding to the
eigenvalue $n$ is simply $z^n$. In other words, $L=z\partial_z$. We  
remark that the expectation value of $r^2=|z|^2$ in this state is $n+1$.

For a system of $N$ bosons, the appropriate wave functions are
analytic and symmetric functions of the bosons coordinates
$z_1,\dots,z_N$. The Hilbert space is thus $\Hh_N=\otimes^N_{\rm symm} 
\Hh$. The kinetic energy is simply $\omega$ times the total
angular momentum.

In addition to the kinetic energy, there is pairwise interaction among
the bosons. It is assumed to be short range compared to any other
characteristic length in the system, and can be modeled by $g 
\sum_{1\leq i<j\leq N}
\delta(z_i-z_j)$ for some coupling constant $g>0$. Physically, this
$g$ is proportional to $a\omega_3^{1/2}$ where $a$ is the scattering  
length of the three-dimensional interaction potential. On the full, original,   
Hilbert space $\otimes^N_{\rm symm} L^2(\mathbb R^2)$, a 
$\delta$-function
as a repulsive interaction potential is meaningless. On the
subspace $\Hh_N$ the matrix elements of $\delta(z_i-z_j)$ make perfect
sense,
however, and define a bounded operator $\delta_{ij}$. Using the  
analyticity of the wave functions this operator
is easily shown to act as
\begin{equation}\label{def12}
\left(\delta_{12} \psi\right)(z_1,z_2) = \frac 1{2\pi}  \psi((z_1+z_2)/
2,(z_1+z_2)/2)\,,
\end{equation}
which takes analytic functions into analytic functions. Its matrix
elements in a two-particle function $\psi(z_1,z_2)$ are
\begin{equation}\label{deltamatrix}
\langle\psi|\delta_{12}|\psi\rangle =  \int_\C |\psi(z,z)|^2 \e{-2|z|
^2} dz\,.
\end{equation}

The dimensional reduction from three to two dimensions for the
$N$-body problem and the restriction to the LLL is, of course, only
reasonable if the interaction energy per particle is much less than
the energy gap $2\hbar\omega_\perp$ between Landau levels and the gap
$\hbar\omega_3$ for the motion in the $x_3$-direction. For a dilute
gas the interaction energy per particle is of the order $a\rho$ where
$\rho$ is the average three-dimensional density
\cite{LY1998}. Provided $Ng/\omega$ is not small we can estimate
$\rho$ by noting that the effective radius $R$ of the system can be
obtained by equating $a\rho\sim Na/(R^2\omega_3^{-1/2})\sim Ng/R^2$
with the kinetic energy $\omega L\sim \omega R^2$. This gives $R\sim
(Ng/\omega)^{1/4}$ and the condition for the restriction to the LLL
becomes
\begin{equation}\label{LLLcond}
(Ng\omega)^{1/2}\ll\min\{\omega_\perp,\omega_3\}.
\end{equation}

The physics of rapidly rotating ultracold Bose gases close to the LLL
regime has been the subject of many theoretical and experimental
investigations in recent years, starting with the papers \cite{wgs, jk, sw, Ho, CWG}. The recent reviews \cite{BDZ,KT,Fe,Coo} contain  extensive lists of
references on this subject. On the experimental side we mention in
particular the papers \cite{ECHSC,SCEMC,CHE} that
report on experiments with rotational frequencies exceeding 99\% of
the trap frequency.

\section{Model and Main Result}

The discussion in the Introduction leads us to the following
well-known model (see, e.g., \cite{wgs,BDZ,Coo}) for $N$ bosons with repulsive
short-range pairwise interactions:
\begin{equation}\label{hamiltonian}
H= \omega \sum_{i=1}^N L_i + g \sum_{1\leq i<j\leq N} \delta_{ij} \,.
\end{equation}
It acts on analytic and symmetric functions of $N$ variables $z_i\in
\C$. The angular momentum operators are $L_i = z_i \partial_{z_i}$,
and $\delta_{ij}$ acts as in (\ref{def12}).  The parameters $\omega$
and $g$ are assumed to be positive. The rigorous derivation of
this model from the 3D Schr\"odinger equation for particles interacting
with short, but finite  range potentials will be presented
elsewhere \cite{LeS}.

Note that the two terms in $H$ commute with each other, and hence can
be diagonalized simultaneously. 
An exact eigenfunction with total angular momentum
$L_{\rm tot} = \sum_{i=1}^N L_i = N(N-1)$ is the Laughlin state
$\prod_{i<j}(z_i-z_j)^2$ for which the interaction vanishes, of
course. For smaller $L$ the interaction is strictly positive. It is
expected that the lowest eigenvalue of the interaction is of the order
$N$ when $L_{\rm tot}$ is of the order $N^2$. The lower boundary of
the joint spectrum of interaction and angular momentum has been called
the {\it yrast} curve \cite{Mo}, a term originating in nuclear
physics. {\it Upper bounds} on this curve can be found by variational
calculations, and a host of trial functions with many interesting properties related to the Fractional Quantum Hall Effect (FQHE) have been employed for this
purpose, see, e.g., \cite{CWG,Coo,vhr,bertsch,nu,yl}. A challenging problem is to establish the missing reliable {\it lower
bounds}.

In this paper we consider the case when $L_{\rm tot}$ is much smaller
than $N^2$, i.e, we investigate one corner of the  asymptotics of the  
yrast region. 
In this case, we shall show that the ground state energy
of $H$ is well approximated (rigorously so in the $N\to\infty$ limit) by the Gross-Pitaevskii energy. The latter
energy is the minimum of the Gross-Pitaevskii functional
\begin{equation}\label{gpfunct}
\E^{\rm GP}(\phi) =  \omega \langle\phi| L | \phi\rangle +  \frac g 2
\int_\C |\phi(z)|^4 \e{-2|z|^2} dz
\end{equation}
over all {\it analytic} functions $\phi$ with
$\langle\phi|\phi\rangle=N$. This can also be viewed as a Hartree
approximation to the many-body system, where one takes the expectation
value of $H$ with a simple product function $\prod_{i=1}^N \phi(z_i)$
(and ignores a factor $(N-1)/N$). The minimization problem for (\ref{gpfunct}) and the properties of the minimizers, in particular their vortex structure, have been studied in some detail by many authors including \cite{Ho, WBP, B, ABD, ABN}.

To state our result precisely, let $E_0(N,\omega,g)$ be the ground state energy of $H
$, and let $E^{\rm GP}(N,\omega,g)$ be the Gross-Pitaevskii energy. We
will find a positive, finite constant $C$ such that
\begin{align}\nonumber 
E^{\rm GP}(N,\omega,g) & \geq E_0(N,\omega,g) \\ & \geq E^{\rm GP}(N,
\omega,g) \left[ 1 - C \left(\frac {g}{N\omega}\right)^{1/10}\right]\label{mainresult}
\end{align}
for all $N$, $\omega$ and $g$ such that $gN/\omega$ is bounded below  
by some (arbitrary) fixed constant. More precisely, for any $c>0$  
there exists a $C<\infty$ such that (\ref{mainresult}) holds if $gN/ 
\omega\geq c$.
In particular, this implies that the ratio of $E_0$ and $E^{\rm GP}$
is close to one if $g\ll N\omega$ and $g \gtrsim N^{-1}\omega$.

Note that, by simple scaling, $E^{\rm GP}(N,\omega,g)= N \omega E^{\rm
    GP}(1,1,gN/\omega)$. Hence the contrary case of small $gN/\omega$ is not  
particularly interesting; as $gN/\omega\to 0$ one obtains a 
non-interacting gas. On the other hand, if $gN/\omega$ is not small the
kinetic and interaction energy are of the same  order of magnitude, which, according to the previous back-of-the-envelope analysis, implies $E^{\rm GP}(N,\omega,g)\sim N\sqrt{Ng\omega}$. 
Hence the total angular momentum, which is obtained by taking
the derivative of the energy with respect to $\omega$, is of the order
$N \sqrt{Ng/\omega}$, which is much less than $N^2$ if and only if $g
\ll N
\omega$. Hence our bound (\ref{mainresult}) covers the whole parameter
regime $L_{\rm tot}\ll N^2$.

We note also that in terms of the {\it filling factor} $\nu=N^2/
(2L_{\rm tot})$ \cite{CWG} the parameter regime $g/(N\omega)\ll 1$
corresponds to $\nu\gg 1$. In contrast, the Laughlin wave function has
filling factor $\nu=1/2$ for $N$ large. Between these two extremes
rich physics related to the FQHE is expected
\cite{Coo}, but this regime is apparently still out of experimental
reach.

\bigskip Before giving the proof of (\ref{mainresult}), we shall
discuss some of its implications for the yrast line. Recall that the
yrast energy $I_0(L)$ is defined as the ground state energy of
$\sum_{i<j}\delta_{ij}$ in the sector of total angular momentum
$L$. What (\ref{mainresult}) says is that
\begin{equation}\label{yr1}
 \omega L + g\, I_0(L) \geq E^{\rm GP}(N,\omega,g) \left[ 1 - C \left(\frac {g}{N\omega}\right)^{1/10}\right]
\end{equation}
for any $L\geq 0$. 
As $Ng/\omega$ gets large, it is expected that 
\begin{equation}\label{assu}
E^{\rm GP}(N,g,\omega) \approx b\, N \sqrt{Ng\omega}
\end{equation}
to leading order, with $b\approx 0.57$. An {\it upper} bound with this
value of $b$ was actually derived in \cite{ab}, but a lower bound with
the same $b$ is still
open. We remark that it is easy to derive a {\it lower} bound 
with $b= \sqrt{8/9\pi} \approx 0.53$. This follows using $\langle
\phi| L|\phi\rangle = \int |\phi(z)|^2 (|z|^2-1) \e{-|z|^2}\dd z$ and
minimizing $\int (\omega |z|^2 |\phi(z)|^2\e{-|z|^2} + (g/2) |\phi(z)|^4
\e{-2|z|^2}) \dd z$ over all $|\phi|^2$, dropping the LLL condition.
Thus, $0.53\leq b \leq 0.57$.  
Using our bound (\ref{yr1}) we deduce from (\ref{assu}) that
$$ 
I_0(L) \gtrsim \frac{b^2}4 \frac{N^3}{L} \quad \text{for $N\ll L
\ll N^2$}\,.  
$$

For $2\leq L\leq N$ it is well known that $I_0(L) = N (4\pi)^{-1} ( N
- 1 -L/2)$ \cite{pap,bertsch,jk,sw,hv}. Our bound (\ref{assu}) reproduces
this result for large $N$ since, as we shall prove below,
\begin{equation}\label{yr2}
E^{\rm GP}(N,\omega,g) = \frac { N^2 g}{4 \pi} \quad \text { if\quad $g\leq \frac{8\pi \omega}{N}$\,,}
\end{equation}
which is just the GP energy of the constant function.  In particular,
the zero angular momentum state is a minimizer of the GP functional
for $gN\leq 8\pi \omega$.  Eqs. (\ref{yr1}) and (\ref{yr2}) imply that
\begin{equation}\label{yrast}
I_0(L) \geq  \frac{N}{4\pi} \left( N  \left[ 1 - C \left(\frac {8\pi}{N^2}\right)^{1/10}\right] - \frac 12 L\right) 
\end{equation}
for all $L\geq 0$. 
In order to prove (\ref{yr2}), we note that the GP energy (\ref{gpfunct}) equals a certain two-particle energy, namely  
$$
\E^{\rm GP}(\phi) = \frac 12  \left\langle \phi\otimes \phi \left| \frac \omega N( L_1 + L_2)  + g\, \delta_{12} \right| \phi\otimes\phi \right\rangle \,.
$$ 
The operator $\delta_{12}$ commutes with $L_1+L_2$, and satisfies $\delta_{12}^2 = (2\pi)^{-1} \delta_{12}$. That is, $2\pi\delta_{12}$ is a projection operator. We claim that 
\begin{equation}\label{yr3}
 \left\langle \phi\otimes \phi \left|  L_1 + L_2 \right| \phi\otimes\phi \right\rangle \geq 4  \left\langle \phi\otimes \phi \left| 1 - 2\pi \delta_{12} \right| \phi\otimes\phi \right\rangle
\end{equation}
for any $\phi\in \Hh$. This clearly implies (\ref{yr2}).  To see
(\ref{yr3}), we can decompose the two-particle function
$\phi(z_1)\phi(z_2)$ into a sum $\sum_{n\geq 0} \psi_n(z_1,z_2) $ of
functions of given total angular momentum $n$. For the terms with
$n\geq 4$ the bound (\ref{yr3}) certainly holds. Also for $n=0$ and
$n=1$ it holds, since $1-2\pi\delta_{12} = 0$ in this case. We thus
have to consider only the cases $n=2$ and $n=3$. On the subspace with
$n=2$, $(1-2\pi\delta_{12})\phi\otimes\phi$ is proportional to the
normalized function $\eta_2(z_1,z_2) = (z_1-z_2)^2 / \sqrt{8}$. It is
straightforward to check that
$|\langle\phi\otimes\phi|\eta_2\rangle|\leq \|\psi_2\|/\sqrt{2}$ which
implies the desired result. Similarly, on the subspace with $n=3$,
$(1-2\pi\delta_{12})\phi\otimes\phi$ is proportional to
$\eta_3(z_1,z_2) = (z_1-z_2)^2(z_1+z_2) / \sqrt{16}$. One checks that
$|\langle\phi\otimes\phi|\eta_3\rangle|\leq \|\psi_3\| \sqrt{3/4}$,
which is what is needed.

\bigskip
Everything now depends on the proof of (\ref{mainresult}) and the most of
the rest of this paper is devoted to this task.  The first inequality in
(\ref{mainresult}) follows easily by taking product wave functions as
trial wave functions, as mentioned earlier. Hence it remains to show
the second inequality in (\ref{mainresult}), which is the lower bound
on $E_0$.

For the proof we shall employ the technique of coherent states and
c-number substitutions \cite{klauder, LSY} that was used in \cite{LS} to
solve a related problem, namely to derive the Gross-Pitaevskii equation
from the three-dimensional many-body Hamiltonian of a rotating Bose gas
away from the critical rotational frequency. This was done in the Gross
Pitaevskii limit where both $\omega$ and $Ng$ are fixed and order unity.
The situation discussed in the present paper is partly simpler than
that in \cite{LS}, where the interaction was described by an arbitrary
repulsive potential of short range instead of a contact interaction. On
the other hand we now have to face new problems. In the
present setting the coupling parameter $gN/\omega$ can vary with $N$
(as long as it is $\ll N^2$) and this fact requires considerably more
delicate estimates on matrix elements of the interaction between
two-particle states than in \cite{LS}. In particular, our bound is
asymptotically exact in the whole "Thomas--Fermi" regime. In order to
make the proof more transparent it will be divided into seven steps.

\section{Derivation of the main inequality (\ref{mainresult})} 

\subsection{Step 1}

As a first step towards deriving  a lower bound on the ground state  
energy of $H$,
we  consider a slightly bigger Hamiltonian $H'$ which is
constructed in the following way. The reason for introducing $H'$ is that the additional small positive term is needed to compensate negative terms that will occur at a later stage of the proof.

For integers $J$, let
\begin{equation}\label{defchi}
\chi_J(L) = \begin{cases}
0 & \text{for $L\leq J$} \\
\exp\left[ - \tfrac 18 \big(\sqrt{L}-\sqrt{J+1}\big)^2\right] &  
\text{for $L \geq J+1$.} \end{cases}
\end{equation}

Pick an integer $J_0\geq 1$. We first show that there exists a $J$  
with $J_0\leq J < 2J_0$ such that
\begin{equation}\label{ba}
\left\langle \sum_{i=1}^N \chi_J(L_i) \right\rangle \leq 2^{3/2}
(1+4  \sqrt\pi)\, \frac{E_0(N,\omega,g)}{\omega J^{3/2}} \,,
\end{equation}
where $\langle\,\cdot\,\rangle$ denote the expectation value in the  
ground state of $H$.  To see this, note that
\begin{align*}
\sum_{J\geq 0} \chi_J(L) &\leq  1 + 2 \int_0^{\sqrt{L}} \exp\left[ -  
\tfrac 18 \big( \sqrt L -s\big)^2\right] s\, ds \\ &\leq 1 + 4 \sqrt { \pi  
L}  \,.
\end{align*}
In particular, since $\chi_J(0)=0$, we have $\sum_J \chi_J(L) \leq  
(1+4\sqrt \pi)\sqrt{L}$ for all integers $L\geq 0$. This implies that
$$
\frac 1{J_0} \sum_{J=J_0}^{2J_0-1} \chi_J(L) \leq (1+4\sqrt\pi) 
\frac{\sqrt{L}}{J_0}\leq  
2^{3/2}
(1+4\sqrt\pi)\frac{L}{J^{3/2}}\,,
$$
where we have used that $L\geq J_0$ in order that the left side be 
non-zero, and $J_0\geq J/2$.
The expectation value of $\sum_{i=1}^N L_i$ is bounded by the total  
energy divided by $\omega$. Hence the bound (\ref{ba}) holds on  
average for $J_0\leq J<2J_0$ and must thus hold for at least one such  
$J$.

We can now pick an $\eta>0$ and consider the modified Hamiltonian
$$
H' = H + \eta J^{3/2} \omega  \sum_{i=1}^N \chi_J(L_i) \,.
$$
By (\ref{ba}) its ground state energy is bounded from above by $E_0 ( 1+ b\,
\eta)$, with $b=2^{3/2} 
(1+4\sqrt\pi)$.  In other words, $E_0$ is  
bounded from below by the ground
state energy of $H'$ divided by $(1+b\, \eta)$. We choose $\eta = (g/N 
\omega)^{1/10}$.

\subsection{Step 2}

In order to derive a lower bound on $H'$, we shall first extend it to
Fock space in the usual way. We do this in order to be able to utilize
coherent states and the lower and upper symbols of the extended operator.

Let $\varphi_j(z) = z^j (\pi j!)^{-1/2}$ denote the normalized
eigenfunctions of the angular momentum operator $L$ with eigenvalue
$j\in \N$. Let $\ad{i}$ and $\an{i}$ denote the corresponding creation
and annihilation operators.  On Fock space, consider the operator
\begin{align}\nonumber\label{hfock}
   \Ham & = \omega \sum_{i\geq 0} \left[ i + \eta J^{3/2} \chi_J(i)
   \right] \ad{i} \an{i}  \\ \nonumber & \quad + \frac g2 \sum_{ijk} \langle
   \varphi_i\otimes\varphi_j| \delta|
   \varphi_k\otimes\varphi_{i+j-k}\rangle \ad{i}\ad{j}\an{k}\an{i+j-k} \\ & \quad + \mu \left( \sum\nolimits_{i\geq 0} \ad{i}
     \an{i} - N \right)^2
\end{align}
for some $\mu>0$. For simplicity we denote $\delta_{12}$ simply by
$\delta$. Note that because of conservation of angular momentum all
other matrix elements of $\delta$ vanish.  The Hamiltonian $H'$ can be
viewed as the restriction of $\Ham$ to the subspace containing exactly
$N$ particles. The value of $\mu$ is irrelevant, since the term
multiplying $\mu$ vanishes in this subspace. In particular, the ground
state energy of $H'$ is bounded from below by the ground state energy
of $\Ham$, for any value of $\mu$.

We introduce coherent states for all angular momentum states up to
$J$.  That is, in Eq.\ (\ref{hfock}) we normal-order all the $\ad{i}$ and $\an{i}$ in the
usual way (which is only relevant for the term multiplying $\mu$ since
all other terms are already normal ordered) and then replace all
$\an{i}$ by complex numbers $\zeta_i$ and all $\ad{i}$ by the conjugate numbers 
$\zeta_i^*$ for $0\leq i\leq
J$.  The resulting operator on  
the Fock space generated by the 
modes $>J$ is called the {\it lower
symbol} of $\Ham$ and will be denoted by $h_\ell(\vec\zeta)$, where
$\vec\zeta$ stands for $(\zeta_0,\dots,\zeta_J)$. Note that
$h_\ell(\vec\zeta)$ does not conserve the particle number, which
explains why it was necessary to embed our $N$-particle Hilbert space
into Fock space.

The ground state energy of $\Ham$ is bounded from below by the ground
state energy of the {\it upper symbol} $h_u(\vec\zeta)$ when minimized over
all parameters $\vec\zeta$.  The upper symbol is given in terms of the
lower symbol as \cite{klauder}
\begin{equation}\label{exp}
h_u(\vec\zeta) = \e{-\sum_{i=0}^J \partial_{\zeta_i}  
\partial_{\zeta_i^*}} h_\ell(\vec \zeta) =  h_\ell(\vec\zeta) +  
U_1(\vec\zeta) + U_2(\vec\zeta)\,,
\end{equation}
where $U_1(\vec\zeta) =-  \sum_{i=0}^J \partial_{\zeta_i}\partial_{\zeta_i^*}
   h_\ell(\vec\zeta)$ equals 
\begin{align}\nn
   &  - \omega \sum_{i\leq J} i - 2 g
   \sum_{0\leq i\leq J} \langle \phi_\zeta\otimes \varphi_i| \delta|
   \phi_\zeta\otimes \varphi_i\rangle \\ \nn &  - 2g \sum_{0<i\leq
     J} \sum_{k>J} \langle\varphi_k\otimes \varphi_i| \delta|
   \varphi_k\otimes \varphi_i\rangle \ad{k}\an{k} \\ \nn &  - \mu
   \left[ \left( 2 \|\phi_\zeta\|^2 -2 N + 2N^> + 1 \right) (J+1) + 2
     \|\phi_\zeta\|^2 \right]
\end{align}
and
\begin{align*}
   U_2(\vec\zeta) & = \frac 12
   \left(\sum_{i=0}^J \partial_{\zeta_i}\partial_{\zeta_i^*} \right)^2
   h_\ell(\vec\zeta) \\ \nn & = g \sum_{0\leq i,j\leq J}\langle
   \varphi_i\otimes \varphi_j| \delta |\varphi_i\otimes\varphi_j\rangle
   + \mu (J+1)(J+2) \,.
\end{align*}
Here, $\phi_\zeta = \sum_{j=0}^J \zeta_j \varphi_j$, $\|\phi_\zeta\|^2
= \langle \phi_\zeta|\phi_\zeta\rangle = \sum_{i=0}^J |\zeta_j|^2$ and
$N^> = \sum_{k>J} \ad{j}\an{j}$. Note that in a Taylor expansion of
the exponential in (\ref{exp}) only the first three terms contribute
since $h_\ell(\vec\zeta)$ is a polynomial of degree four.

For a lower bound, we can use the fact that $U_2(\vec\zeta)\geq
0$. Moreover, to bound the various terms in $U_1(\vec\zeta)$ we can
use the fact that $\sum_{i\leq J} |\varphi_i(z)|^2 \leq \pi^{-1}
e^{|z|^2}$. It is then easy to see that $U_1(\vec\zeta)$ is bounded
from below as
\begin{align}\nonumber
   U_1(\vec\zeta) &\geq - \frac \omega 2 J (J+1) -\left(\frac{2g}\pi  
+2\mu (J+2)\right) \|\phi_\zeta\|^2 \\ & \quad  - 2  N^> \left[ \mu (J+1) + \frac g 
\pi \right] \,. \label{inp}
\end{align}
In order for the first term to be much smaller than the GP energy
$E^{\rm GP}(N,g,\omega)\sim N \sqrt{Ng\omega}$ it is clear that we can
not replace all modes by a c-number but must rather require that $J\ll N
(g/N\omega)^{1/4}$.

\subsection{Step 3}

As we have explained in the previous step, $H'$ is bounded from below
by the minimum over $\vec\zeta$ of the ground state energy of
$h_u(\vec\zeta)$. Suppose the minimum is attained at some
$\vec\zeta_0$, and let $\langle\,\cdot\,\rangle_0$ denote the
corresponding ground state expectation value in the Fock space of the modes $>J$. The next step is to use a simple lower bound on $h_u(\vec\zeta_0)$ to bound the $\eta$-dependent term in $\mathbb H$ involving the modes $>J$.

Let us choose the parameters $\mu$ and $J$ in following way. Recall  
from step 1
that $J_0\leq J<2J_0$, with $J_0$ an arbitrary integer that we are  
free to choose.  We take $J_0=\lfloor  N
(g/N\omega)^{3/10}\rfloor$, where $\lfloor t\rfloor$ denotes the
integer part of a number $t>0$, and $\mu = \omega J/ (4N)$.  The
proportionality constants here are chosen more for convenience than
for optimality. Note that $J_0$ is a big number if $g/(N\omega)\ll 1$  
and $gN/\omega \gtrsim 1$.

  Recall that $\eta = (g/N\omega)^{1/10}$.  We claim
that, for small $g/N\omega$,
\begin{equation}\label{c1}
   \eta \left\langle \sum_{i>J}  \chi_J(i)  \ad{i} \an{i} \right 
\rangle_0 \leq 2\, \frac{E^{\rm GP}(N,\omega,g)}{\omega J^{3/2}}\,.
\end{equation}
This bound is similar to (\ref{ba}) except for the additional factor
$\eta$ on the left side and $E^{\rm GP}$ replacing $E_0$ on the right side.
In order to prove (\ref{c1}), we can use the positivity of the
interaction to obtain the following lower bound on
$h_u(\vec\zeta)$. The lower symbol $h_\ell(\zeta)$ is bounded from
below as
\begin{align}\nn
 h_\ell(\vec\zeta) 
& \geq \eta \sum_{i>J} \chi_J(i) \ad{i} \an{i} +
   \omega (J+1) N^> \\ \nn &\quad + \mu \biggl[ \left( \|\phi_\zeta\|^2 -
       N\right)^2 +2 N^> (\|\phi_\zeta\|^2 -N) \\ & \qquad\qquad + (N^>)^2 +
     \|\phi_\zeta\|^2 \biggl] \,. \label{c11}
\end{align}
In particular, using (\ref{inp}),
\begin{align}\nn
   h_u(\vec\zeta) & \geq \eta \omega J^{3/2} \sum_{i>J} \chi_J(i)
   \ad{i} \an{i}  + \mu \left( \|\phi_\zeta\|^2 -
     N\right)^2\\ \nn &\quad + N^> \left[ (J+1) (\omega -2 \mu) - \frac{2 g}\pi -2
     \mu N\right] \\ \nn & \quad - N \left(\frac{2g}\pi +2\mu (J 
+2)\right) -
   \frac \omega 2 J (J+1) \\ \nn &\quad -\left(\frac{2g}\pi +2\mu (J+2)\right) ( \|\phi_\zeta\| 
^2 -N)\,.
\end{align}
With the aid of Schwarz's inequality, we obtain
\begin{align}\nn
   h_u(\vec\zeta) & \geq \eta \omega J^{3/2} \sum_{i>J} \chi_J(i)
   \ad{i} \an{i} \\ \nn &\quad + N^> \left[ (J+1) (\omega -2 \mu) - \frac{2 g}\pi -2  
\mu N\right] \\ \nn
   & \quad - N \left(\frac{2g}\pi +2\mu (J+2)\right) \\ &\quad  - \frac \omega 2 J
   (J+1) - \frac{\left(\frac{2g}\pi +2\mu
       (J+2)\right)^2}{4\mu}\,. \label{mt}
\end{align}
For our choice of $\mu$ and $J$, $\mu\ll \omega$ and $g\ll \omega J$
for small $g/N\omega$. Moreover, $\mu N = \omega J/4$. Hence the
second term on the right side of (\ref{mt}) is positive for small
$g/N\omega$. Moreover, since $E^{\rm GP}(N,\omega,g) \sim N
\sqrt{N\omega g}$ for $N g\gtrsim \omega$, as remarked earlier, all
the terms in the last line of (\ref{mt}) are much smaller than $E^{\rm
   GP}(N,\omega,g)$ if $g\ll N\omega$ and, in particular, are bounded  
from below by
$- E^{\rm GP}/2$. Since the ground state energy of $h_u(\vec\zeta_0)$ is
bounded above by $E_0(N,\omega, g)(1+b\, \eta) \leq 1.5\, E^{\rm
   GP}(N,\omega, g)$, Inequality (\ref{c1}) follows.

\subsection{Step 4}
We now give a more refined lower bound on $h_\ell(\vec\zeta)$. Instead  
of dropping all the interaction terms, as we did in the previous step  
in order to obtain the bound (\ref{c11}), we use the fact that
\begin{multline}\label{dpq}
\delta \geq P_J\otimes P_J \delta P_J\otimes P_J + P_J\otimes P_J  
\delta (P_J \otimes Q_J + Q_J\otimes P_J) \\ +  (P_J \otimes Q_J + Q_J 
\otimes P_J)\delta P_J\otimes P_J\,,
\end{multline}
where $Q_J=\id-P_J$ and $P_J$ denotes the projection onto the $J+1$
dimensional subspace of $\Hh$ spanned by eigenfunctions of the angular
momentum with $L\leq J$. Inequality (\ref{dpq}) is a simple Schwarz  
inequality and follows from
positivity of $\delta$ and the fact that
$P_J\otimes P_J \delta Q_J\otimes Q_J = 0$. Using (\ref{dpq}), we see  
that
$h_\ell(\vec\zeta)$ is bounded from below by
\begin{align}\nn
h_\ell(\vec\zeta)&  \geq \E^{\rm GP}(\phi_\zeta)
+ \omega \sum_{j>J} j \ad{j} \an{j}  \\ \nn & 
\quad + \mu \biggl[ \left( \|\phi_\zeta\|^2 - N\right)^2 +2 N^> (\|\phi_ 
\zeta\|^2 -N) \\ \nn &\qquad \qquad + (N^>)^2 + \|\phi_\zeta\|^2 \biggl] \\ \nn &\quad + g  
\sum_{k>J} \langle \phi_\zeta\otimes \phi_\zeta| \delta| \phi_\zeta 
\otimes \varphi_k\rangle \an{k} + \text{h.c.} \,.
\end{align}
For the remaining terms in the upper symbol $h_u(\vec\zeta)$ we  
proceed as in the previous step. We retain $\mu(\|\phi_\zeta\|^2- 
N)^2/2$ for later use, however, and arrive at
\begin{align}\nn
h_u(\vec\zeta)&  \geq \E^{\rm GP}(\phi_\zeta) + \frac \mu2 \left( \| 
\phi_\zeta\|^2 - N\right)^2 \\ \nn &\quad + g \sum_{k>J} \langle \phi_\zeta\otimes  
\phi_\zeta| \delta |\phi_\zeta\otimes \varphi_k\rangle  \an{k} +  
\text{h.c.}  \\ \nn  &\quad  - N \left(\frac{2g}\pi +2\mu (J 
+2)\right) - \frac \omega 2 J (J+1) \\ &\quad  -
   \frac{\left(\frac{2g}\pi +2\mu (J+2)\right)^2}{2\mu} \,.\label{hh2}
\end{align}

The right side of (\ref{hh2}) contains the desired quantity $\E^{\rm
   GP}(\phi_\zeta)$. The second term guarantees that $\|\phi_\zeta\|^2$
is close to $N$. All the terms in the last two lines are small compared to
$E^{\rm GP}(N,\omega,g)$ for our choice of $\mu$ and $J$. Hence we are
left with giving a bound on the third term in (\ref{hh2}), which is
linear in creation and annihilation operators. It is this  bound that requires most effort; it will be completed in the remaining three steps.

\subsection{Step 5}

A simple Schwarz inequality shows that, for any sequence of complex  
numbers $c_k$ and positive numbers $e_k$,
\begin{equation}\label{this1}
  \sum_k \left( c_k \an{k} + c_k^* \ad{k} \right)  \leq  \sum_k \frac{| 
c_k|^2}{e_k} + \sum_k e_k \ad{k}\an{k}  \,.
\end{equation}
We apply this to our case, where $c_k = - \langle\phi_\zeta\otimes  
\phi_\zeta| \delta |\phi_\zeta \otimes \varphi_k\rangle$ for $k>J$,  
and $c_k=0$ otherwise. We pick some $\kappa>0$ and choose $e_k =  
\kappa^{-1} \chi_J(k)$ for $k>J$, with $\chi_J(k)$ given in  
(\ref{defchi}).
Then
\begin{equation}\label{twt}
\sum_k \frac{|c_k|^2}{e_k} = \kappa \langle\phi_\zeta| \rho_\zeta  
\chi_J^{-1}(L) \rho_\zeta | \phi_\zeta\rangle\,,
\end{equation}
where $\rho_\zeta$ denotes the multiplication operator $\rho_\zeta(z)  
= |\phi_\zeta(z)|^2 \e{-|z|^2}$ and
$$
\chi_J^{-1}(L) \equiv \sum_{k=J+1}^{\infty} \frac 1{\chi_J(k)} | 
\varphi_k\rangle\langle\varphi_k|\,.$$
(For simplicity, we abuse the notation slightly, since $\rho_\zeta$  
does not leave $\Hh$ invariant;  $\langle \varphi_k|\rho_\zeta|\phi_ 
\zeta\rangle$ makes perfectly good sense, however.)

In Step 7 below we shall show that
\begin{equation}\label{key}
\langle\phi_\zeta| \rho_\zeta \chi_J^{-1}(L) \rho_\zeta| \phi_\zeta 
\rangle \leq 6\,  \left[ \int_\C  |\phi_\zeta(z)|^4 \e{-2 |z|^2 } dz  
\right]^{3/2}\,.
\end{equation}
This inequality may seem surprising at first sight. The
function $\rho_\zeta|\phi_\zeta\rangle$ contains angular momenta
up to $L\leq 2J$, and $\chi_J^{-1}(2J)$ grows exponentially with
$J$. As will become clear below, however, $|\langle \varphi_k|\rho_ 
\zeta|\phi_\zeta\rangle|^2$ is exponentially small unless $k$ is  
smaller than roughly $J + \sqrt{J}$.
Since $\chi^{-1}_J(J + \sqrt{J})$ is bounded independently of
$J$, this makes (\ref{key}) plausible.

Altogether, (\ref{this1})--(\ref{key}) imply that
\begin{multline*}
  \sum_k \left( c_k \an{k} + c_k^* \ad{k}\right)   \\ \leq   6\, \kappa  
\left[ \int_\C  |\phi_\zeta(z)|^4 \e{-2 |z|^2 } dz \right]^{3/2} +   
\sum_k e_k \ad{k}\an{k} \,.
\end{multline*}
We have shown in Step~3 above that
$$
\left\langle \sum_k e_k \ad{k}\an{k} \right\rangle_0 \leq  2\, \frac  
{E^{\rm GP}(N,\omega,g)}{\kappa\eta\omega J^{3/2}}
$$
in the ground state of $h_u(\vec\zeta_0)$.
After optimizing over $\kappa$ we get the following lower bound on $- 
\sum_k(c_k\an{k} + c_k^* \ad{k})$ (valid as an expectation value in  
the ground state):
$$
-  4 \sqrt{3}\,  \left[  \int_\C  |\phi_\zeta(z)|^4 \e{-2 |z|^2 } dz  
\right]^{3/4}  \left( \frac {E^{\rm GP}(N,\omega,g)}{\eta\omega  
J^{3/2}}  \right)^{1/2}\,.
$$
Finally, using H\"older's inequality  this expression is bounded from  
below by
$$
- \frac \beta 2  \int_\C  |\phi_\zeta(z)|^4 \e{-2 |z|^2 } dz - 
\frac {6^5} {\beta^3}  \left(\frac {E^{\rm GP}(N,\omega,g)}{\eta 
\omega J^{3/2}}   \right)^{2}
$$
for arbitrary $\beta>0$.

\subsection{Step 6}

The analysis in the preceding steps has shown that
\begin{align}\nn
h_u(\vec\zeta_0) & \geq  (1-\beta) \E^{\rm GP}(\phi_{\zeta_0}) + \frac \mu 2  
\left( \|\phi_{\zeta_0}\|^2 - N\right)^2 \\ \nn &\quad 
  - N \left(\frac {2g}\pi +2\mu (J+2)\right)  -
   \frac{\left(\frac{2g}\pi +2\mu (J+2)\right)^2}{2\mu}  \\ \nn &\quad - \frac  
\omega 2 J (J+1)  -   g  
 \frac {6^5}{\beta^{3}} \left(\frac {E^{\rm GP}(N,\omega,g)} 
{\eta\omega J^{3/2}}   \right)^{2} \,.\\ \label{ana}
\end{align}
As mentioned above, we choose $\mu = \omega J/(4N)$ and $J\sim N (g/N 
\omega)^{3/10}$. Moreover, we make the choice $\eta =\beta = (g/N 
\omega)^{1/10}$. Then all the error terms are small compared to  
$E^{\rm GP}(N,\omega,g)$, namely they are of the order $(g/N 
\omega)^{1/10} E^{\rm GP}$.

Since the ground state energy of $h(\vec\zeta_0)$ is bounded above by  
$E^{\rm GP}(N,\omega,g)(1+b\, \eta) \leq 1.5\, E^{\rm GP}(N,\omega,g)$ 
and all the error terms in (\ref{ana}) are bounded by $E^{\rm GP}/2$  
for small $g/(N\omega)$, we see that
$$
\left| \|\phi_\zeta\|^2 - N\right| \leq \left( \frac{2\, E^{\rm GP}(N, 
\omega,g)}{\mu}\right)^{1/2} \sim  N \left( \frac g{N\omega} 
\right)^{1/10}
$$
for $\vec\zeta=\vec\zeta_0$ and hence
\begin{multline*}
  (1-\beta) \E^{\rm GP}(\phi_\zeta) + \frac \mu2 \left( \|\phi_\zeta\| 
^2 - N\right)^2\\ \geq E^{\rm GP}(N,\omega,g) \left(1- C \left( \frac g{N 
\omega}\right)^{1/10}\right)
\end{multline*}
for some constant $C>0$.
This yields the lower bound in (\ref{mainresult}).

\subsection{Step 7}

The only thing left to do is to prove Inequality~(\ref{key}). For  
convenience, we shall introduce the notation
$$
\|\phi\|_p = \left( \int_\C |\phi(z)|^p \e{-(p/2)|z|^2} dz\right)^{1/p}
$$
for $1\leq p<\infty$.
We can bound
$$
\langle \phi_\zeta| \rho_\zeta \chi_J^{-1}(L) \rho_\zeta| \phi_\zeta 
\rangle \leq  \|\phi_\zeta\|_4^2 \, \| \sqrt{\rho_\zeta} \chi_J^{-1} 
(L) \sqrt{\rho_\zeta}\| \,.
$$
The latter norm is an operator norm on $L^2(\C,\e{-|z|^2}dz)$ (defined as the maximum expectation value), and  
equals the operator norm of $\sqrt{\chi_J^{-1}(L)}\rho_\zeta \sqrt{\chi_J^{-1} 
(L)}$ on $\Hh$.

Next, we derive a pointwise bound on $\rho_\zeta(z)=|\phi_\zeta(z)|^2
\e{-|z|^2}$. Let $P_J(z,z')=\sum_{j=0}^J \varphi_j(z) \varphi_j(z')^*$
denote the integral kernel of the projection onto the subspace of $\Hh$
with angular momentum $L\leq J$. With the aid of H\"older's inequality,
\begin{align*}
|\phi_\zeta(z)| & = \left| \int_\C P_J(z,z') \phi_\zeta(z') \e{-|z'| 
^2}dz' \right| \\ & \leq \|\phi_\zeta\|_4 \|P_J(z,\,\cdot\,)\|_{4/3}\,.
\end{align*}
Another application of H\"older's inequality yields
$$
  \|P_J(z,\,\cdot\,)\|_{4/3} \leq  \|P_J(z,\,\cdot\,)\|_{2}^{(4-3p)/ 
(4-2p)}  \|P_J(z,\,\cdot\,)\|_{p}^{p/(4-2p)}
$$
for any $1\leq p\leq 4/3$.

The $L^2$ norm of $P_J(z,\,\cdot\,)$ is given by
$$
  \|P_j(z,\,\cdot\,)\|_{2}^2 = \sum_{j=0}^J |\varphi_j(z)|^2\,.
$$
The $L^p$ norms with $p\neq 2$ are not given by simple expressions,
however. In the following, we will show that they can be bounded above
by a quantity independent of $J$. More precisely, for any $1<p<\infty 
$, we shall
show that
$$
\|P_J(z,\,\cdot\,)\|_{p} \leq \frac{(2\pi/p)^{1/p}}{\pi\sin(\pi/p)}  
\e{|z|^2/2}
$$
independently of $J$. In order to see this, note that, for fixed $z$
and $|z'|$, the function $\theta \mapsto P_J(z,|z'|\e{i\theta})$ is
obtained from $P_\infty(z,|z'|\e{i\theta})$ by restricting the Fourier
components to $0\leq j\leq J$.  This restriction is a bounded
operation (uniformly in $J$) on $L^p(\mathbb{T})$ for $1<p<\infty$
\cite[Ch.~II, Sec.~1]{Katznelson}, and hence $\|P_J(z,\,\cdot\,)\|_{p}$ is bounded by 
a constant times $\|P_\infty(z,\,\cdot\,)\|_{p}$. In fact, an upper
bound on the optimal constant in this inequality is given by the norm
of the Riesz projection (viewed as an operator from $L^p(\mathbb{T} )$
to itself), which is known to equal $1/\sin(\pi/p)$ \cite{Hollenbeck}.

Hence we have the bound
\begin{align}\nn
\|P_J(z,\,\cdot\,)\|_{p} &\leq \frac 1{\sin(\pi/p)} \|P_\infty(z,\,\cdot 
\,)\|_p \\ &= \left[ \frac{(2\pi/p)^{1/p}}{\pi \sin(\pi/p)}\right] \e{|z| 
^2/2}  \,. \label{cp}
\end{align}
The last equality follows from the fact that $P_\infty(z,z')=\pi^{-1}  
\e{-z z'^*}$.

Let $c_p$ denote the constant in square brackets in (\ref{cp}) taken to the power $2p/(4-2p)$.
We have thus shown that, for $1<p\leq 4/3$ and $\alpha(p) = (4-3p)/ 
(4-2p)$
$$
\rho_\zeta(z)  
\leq c_p \|\phi_\zeta\| 
_4^2 \left( \sum\nolimits_{j=0}^J |\varphi_j(z)|^2 \e{-|z|^2}  
\right)^{\alpha(p)}\,.
$$
A possible choice is $p=6/5$, in which case $c_p= (10/3) \sqrt{2} (5/3\pi)^{1/4} \approx 4.02$ and $\alpha(p)=1/4$.
Hence, for any function $|\psi\rangle = \sum_{k>J} d_k |\varphi_k 
\rangle$, we have that
\begin{align}\nn
&\left\langle \psi\left| \sqrt{\chi^{-1}_J(L)}\rho_\zeta  
\sqrt{\chi_J(L)}\right| \psi\right\rangle \\ & \nn \leq 4.02\, \|\phi_\zeta\|_4^2 \sum_{k 
\geq J+1} |d_k|^2 \exp\left[ \tfrac 18 \big(\sqrt{k}-\sqrt{J 
+1}\big)^2\right] \\ &  
\quad \times \int_{\C} | 
\varphi_{k}(z)|^2 \left( \sum\nolimits_{j=0}^J |\varphi_j(z)|^2 \e{-|z| 
^2} \right)^{1/4} \e{-|z|^2} dz\,. \label{expin}
\end{align}
Here we have used the fact that our upper bound on $\rho_\zeta(z)$ is
radial, i.e., depends only on $|z|$, and hence there are no
off-diagonal terms on the right side.  An application of Jensen's
inequality for the concave function $t \mapsto t^{1/4}$ implies that
\begin{align}\nn
&\int_{\C} |\varphi_{k}(z)|^2 \left( \sum\nolimits_{j=0}^J | 
\varphi_j(z)|^2 \e{-|z|^2} \right)^{1/4}\e{-|z|^2} dz \\ \nn &\leq  
\left( \int_{\C} |\varphi_{k}(z)|^2 \e{-|z|^2}\left( \sum 
\nolimits_{j=0}^J |\varphi_j(z)|^2 \e{-|z|^2} \right) dz \right)^{1/4}  
\,.
\end{align}

{}From  Stirling's formula \cite{AS} it follows that $j! \geq (j/{\rm  
e})^j \sqrt{1+j}$ for any $j\geq 0$. Using this bound  it is easy to  
see that
$$
|\varphi_j(z)|^2 \e{-|z|^2} \leq  \frac{1}{\pi \sqrt{1+j}} \e{-(|z|- 
\sqrt{j})^2}\,.
$$
Similarly,
$$
|\varphi_k(z)|^2 \e{-|z|^2} \leq \frac{3^{3/2}}{\pi \sqrt{8\,{\rm e}}}  
\frac{1}{|z|} \e{-(|z|-\sqrt{k})^2}
$$
for $k\geq 1$.
Hence
\begin{align*}\nn
& \int_\C |\varphi_j(z)|^2 |\varphi_k(z)|^2 \e{-2|z|^2} dz \\ &\leq  
\frac{3^{3/2}}{\pi \sqrt{2\,{\rm e}}} \frac 1{\sqrt{1+j}} \int_0^ 
\infty \e{-(x-\sqrt{j})^2} \e{-(x-\sqrt{k})^2} dx \\ \nn &\leq    
\frac{3^{3/2}}{\sqrt{4 \pi\,{\rm e}}} \frac 1{\sqrt{1+j}} \e{- 
(\sqrt{j}-\sqrt{k})^2/2}\,.
\end{align*}
To obtain the last inequality, we simply extended the integral to the  
whole real axis.
The sum over $j$ can be bounded from above by the integral
\begin{align*}\nn
&\sum_{j=0}^J \frac{1}{\sqrt{1+j}} \e{-(\sqrt{j}-\sqrt{k})^2/2} \\ &\leq   
\int_0^{J+1} \e{-(\sqrt s-\sqrt{k})^2/2} \frac{ds}{\sqrt s} \\ & \leq  
2 \int_{-\infty}^0 \e{-(t + \sqrt{J+1}-\sqrt{k})^2/2} dt \\ \nn & \leq  
2\int_{-\infty}^\infty \e{-(t + \sqrt{J+1}-\sqrt{k})^2/2} \e{-  
t(\sqrt{k}-\sqrt{J+1})} dt  \\ \nn & = \sqrt{8\pi} \e{-(\sqrt{k}- 
\sqrt{J+1})^2/2} \,.
\end{align*}
The one-fourth root of this expression cancels exactly the exponential  
factor in (\ref{expin}), and we arrive at (\ref{key}).

\section{Bose-Einstein Condensation} 

The technique employed for the
energy bounds can be used to show that the model (\ref{hamiltonian})
exhibits Bose-Einstein condensation in the ground state if
$N\to\infty$ with $gN/\omega$ fixed, in complete analogy to the proof
of the corresponding result in \cite{LS}. The nonuniqueness of the
minimizers of the GP functional (\ref{gpfunct}) due to breaking of
rotational symmetry makes the precise statement a little complicated,
but in brief the result is as follows: If $\Psi_N$ is a sequence of
ground states of (\ref{hamiltonian}) for $N=1,2,\dots$ and $\gamma_N$
are the corresponding normalized 1-particle density matrices, then any
limit $\gamma$ of the sequence $\gamma_N$ as $N\to\infty$ with the
coupling constant $gN/\omega$ fixed is a convex combination of
projectors onto normalized minimizers of the GP functional
(\ref{gpfunct}) for the given coupling constant. A precise formulation
is given in Theorem 2 in \cite{LS}. A key step in the proof is to
extend the bounds on the ground state energy of $H$ to perturbed
operators $H^{(S)}=H+\sum_{i=1}^NS_i$ where $S$ is a bounded operator
on the 1-particle space $\mathcal H$ and $S_i$ the corresponding
operator on the $i$-th factor in $\otimes_{\rm symm}^N\mathcal
H$. These bounds lead to the inequalities
\begin{equation}
\min_\phi\langle\phi|S|\phi\rangle\leq \Tr S\gamma\leq \max_\phi\langle\phi|S|\phi\rangle
\end{equation}
for any limit $\gamma$ of the sequence of density matrices $\gamma_N$,
where the $\max$ resp.\ $\min$ is taken over all normalized minimizers
$\phi$ of (\ref{gpfunct}) with coupling parameter $gN/\omega$. With
the aid of some arguments from convex analysis one can then conclude
as in \cite{LS} that $\gamma$ has a representation in terms of
projectors $|\phi\rangle\langle\phi|$ onto normalized minimizers
$\phi$ of the GP functional. The appropriate extension of this result
to the Thomas-Fermi limit is still an open problem.

\section{Conclusion}
We analyzed the ground state energy of a rapidly rotating Bose gas in a harmonic trap,
under the usual assumption that the particles are restricted
to the lowest Landau level, that the two-body interaction is a
$\delta$-function, and that the number of particles, $N$, is very
large. For the limiting situation we prove, rigorously, that the
Gross-Pitaevskii energy, appropriately modified to the lowest Landau
level, is exact
 provided $g\ll N\omega$. 
Here, $g$ is the strength of the $\delta$-function interaction,
and $\omega$ is the difference between the trapping and rotation
frequencies.

\section*{Acknowledgments}
We thank Rupert Frank for illuminating discussions. This work was
partially supported by U.S. National Science Foundation grants
PHY-0652854 (E.L.) and PHY-0652356 (R.S.), and the ESF network INSTANS (J.Y.).

\end{document}